\documentclass[aps,prb,superscriptaddress, twocolumn]{revtex4-1}
\usepackage{graphicx}
\usepackage{dcolumn}
\usepackage{bm}
\usepackage{amsmath}

\usepackage{color}
\usepackage{bm,graphicx,hyperref}
\hypersetup{%
  breaklinks = {true},
  citecolor = {blue},
  colorlinks = {true},
  linkcolor = {red},
}

\begin{document}

\title{Bulk-mediated interaction between impurities in 1D atomic chains}

\author{Aleksandr Rodin}\thanks{Corresponding author: aleksandr.rodin@yale-nus.edu.sg}
\affiliation{Yale-NUS College, 16 College Avenue West, 138527, Singapore}
\affiliation{Centre for Advanced 2D
  Materials, National University of
  Singapore, 6 Science Drive 2, 117546, Singapore}

\author{Keian Noori}
\affiliation{Centre for Advanced 2D
  Materials, National University of
  Singapore, 6 Science Drive 2, 117546, Singapore}

\author{Su Ying Quek}
\affiliation{Centre for Advanced 2D
  Materials, National University of
  Singapore, 6 Science Drive 2, 117546, Singapore}
\affiliation{Department of Physics, National University of
  Singapore, 2 Science Drive 3, 117542, Singapore}  

\date{\today}

\begin{abstract}

A combination of numerical and analytical methods is employed to study a one-dimensional chain of identical atoms with adsorbates. We show that the electron-mediated interaction energy between two impurities can change sign and magnitude depending on the adatom-adatom separation, as well as the system doping. By focusing on this simple system, we provide insight into the bulk-mediated interaction for more complex materials.

\end{abstract}

\maketitle

\section{Introduction}
\label{sec:Introduction}

Impurities in a material can interact via their host's electrons when the impurity energy levels hybridize with the electronic states of the bulk. One example is spin-spin interaction, commonly studied using the RKKY formalism, which can lead to (anti-) ferromagnetic ordering. Another is the potential interaction whose nature guides the arrangement of the adatoms by causing them either to cluster or disperse.

With the rise of two-dimensional materials, the interest of the condensed matter community turned to graphene-hosted adatoms. In their pioneering work, the authors of Ref.~\onlinecite{Shytov2009lri} claimed that the nature of the impurity-impurity interaction was determined by the sublattice configuration. The prediction was that atoms on the same sublattice repel while those on the opposite sublattices attract. More recent studies~\citep{LeBohec2014art, Agarwal2019pas}, however, showed that while the sublattice arrangement is important, there are also other factors that play a role in the interaction. It was demonstrated, for example, that the interaction can change sign with the separation between the impurities and their on-site energies. First principles calculations have also shown that adsorbed atoms can attract regardless of the host sublattice.~\citep{GonzalezHerrero2016asc}

Despite the relative simplicity of the graphene lattice, the mathematics involved in calculating the interaction between impurities analytically is substantial and this can obscure the physical picture. This complexity has, in part, led to the conflicting results of earlier publications. In order to gain a better understanding of the physical mechanisms behind the impurity-impurity interaction, it is useful to focus on a system that is more tractable mathematically.

Here, we consider a one-dimensional chain of identical atoms with two adsorbed impurities. By using a combination of analytical and numerical methods, we show that even this simple system  exhibits a rich behavior where the interaction can change sign based on the doping of the system, as well as the separation between impurities. The simplicity of the model allows us to employ a non-perturbative approach to construct an intuitive picture of the processes involved in the interaction which can be generalized to other, more complex, systems.

\section{Model}
\label{sec:Model}

\subsection{Atomic Chain}
\label{sec:Atomic_Chain}

\begin{figure}[h]
	\includegraphics[width = 3.0in]{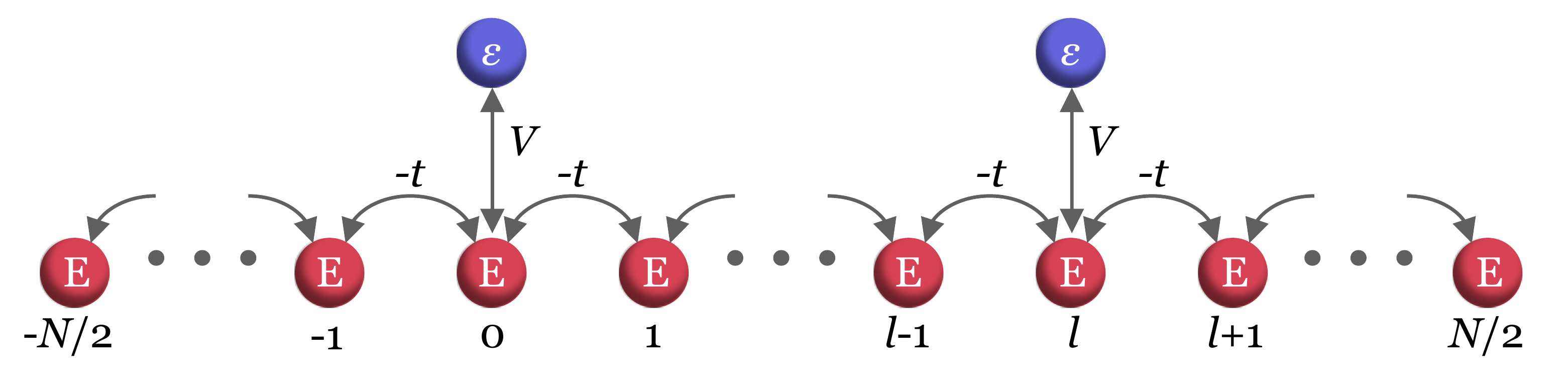}
	\\
	\includegraphics[width = 3.0in]{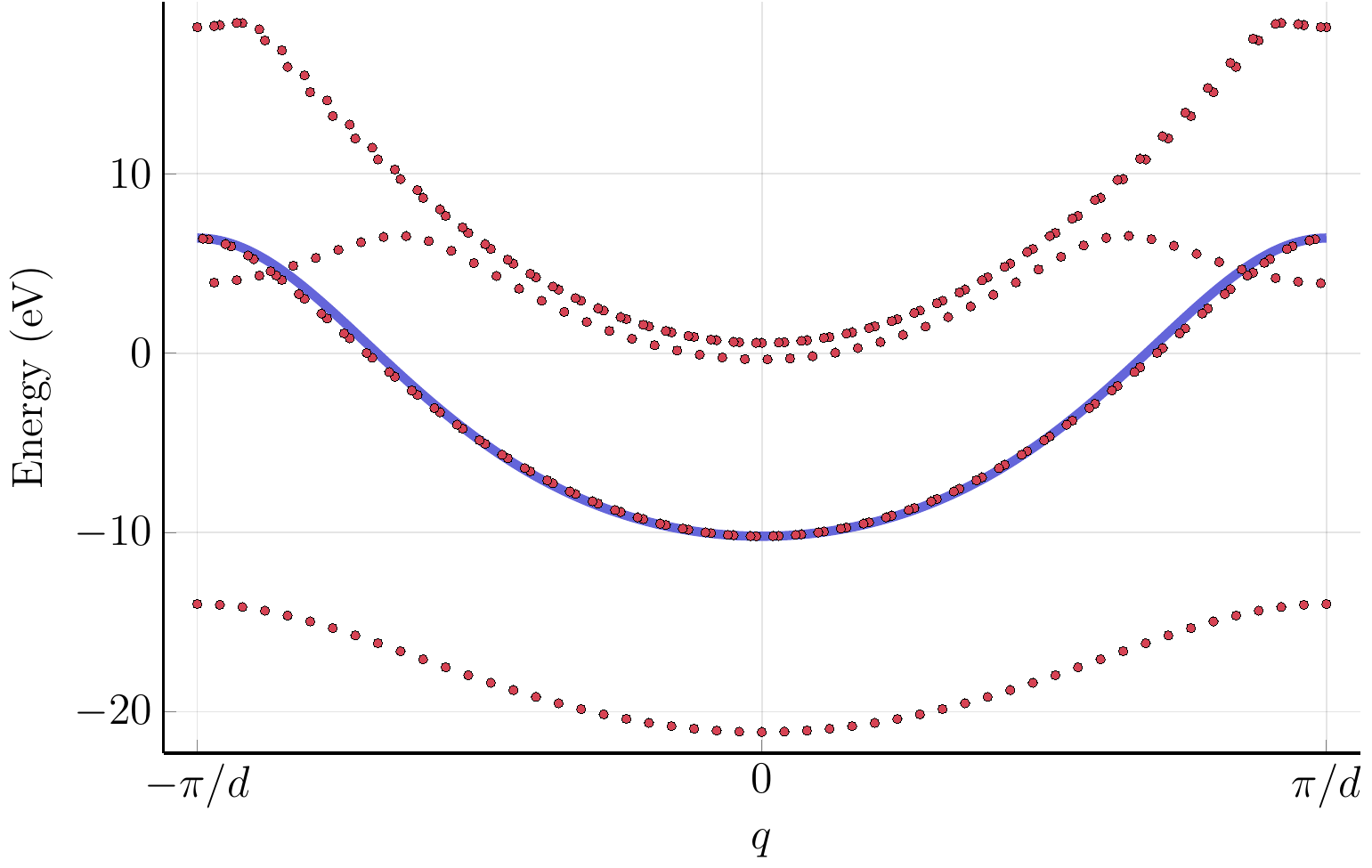}
	\caption{\textbf{Top:} Atomic chain with adsorbed impurities. The numbers under the chain label the atoms, $V$ and $-t$ are the hopping parameters, and the values on the atoms are the site energies. \textbf{Bottom:} \emph{Ab initio} band structure (dots) for an infinite 1D chain of C atoms, and the corresponding tight-binding model for the $\pi$ bands (solid line). The parameters for the tight-binding fit are: $t = 4.49$ eV, $g = 0.1776$, $E = -4.851$ eV.}
	\label{fig:Chain}
\end{figure}

The system in question is composed of two coupled components: the atomic chain and the adsorbed impurities. To model the chain, we use the tight-binding formalism with nearest neighbor hopping for an equally-spaced arrangement of single-orbital atoms, see Fig.~\ref{fig:Chain}. The dispersion for such a system is given by
\begin{equation}
	\mathcal{E}_q = \frac{E-2t\cos\left(qd\right)}{1 + 2g\cos\left(qd\right)}\,,
	\label{eqn:Dispersion}
\end{equation}
where $-t$ is the hopping parameter, $g$ is the orbital overlap between the neighboring atoms, $E$ is the on-site energy, and $d$ is the interatomic distance.

In this work, we compare analytic results with \emph{ab initio} density functional theory (DFT) calculations. Therefore, we need to ensure that Eq.~\eqref{eqn:Dispersion} adequately captures the relevant features of the \emph{ab initio} band structure of our model system.  To this end, we constructed a model 1D chain of C atoms with equal spacing of 1.29 \AA.  

The DFT band structure of the 1D chain is given in Fig.~\ref{fig:Chain}, with the doubly-degenerate band of interest lying between $-10$ and $6.5$ eV.  The two bands are composed of $\pi$-bonded $p$ orbitals perpendicular to the direction of the chain. By fitting Eq.~\eqref{eqn:Dispersion} to the \emph{ab initio} data at $qd = 0$, $\pi/2$, and $\pi$, we obtain $E$, $g$, and $t$ and plot the analytic formula along with the numerical results.

\subsection{Chain-Impurity Coupling}
\label{sec:Chain_Impurity_Coupling}

We assume that each adatom has a single $s$ orbital with energy $\epsilon$. The impurity binds to a $p$ orbital of an individual chain atom with the hybridization energy $V$. Positioning one impurity at the 0th atom in the chain and another at the $l$th atom (see Fig.~\ref{fig:Chain}) yields the following Hamiltonian:
\begin{align}
	H  
	=& \sum_{q\in \mathrm{BZ}} c_q^\dagger\left[ \mathcal{E}_q- \mu\right] c_q
	+ \left(\epsilon - \mu\right)\left(a^\dagger a + b^\dagger b\right)
	\nonumber
	\\
	+& \frac{V}{\sqrt{N}} \sum_{q\in \mathrm{BZ}} c_q^\dagger\left[ a  +  b e^{-iqdl} \right] 
		+ \left[a^\dagger + b^\dagger  e^{iqdl}\right] c_q\,,
		\label{eqn:H}
\end{align}
where $N$ is the number of states in the one-dimensional Brillouin zone. The operators $c_q$ correspond to the chain states, while $a$ and $b$ are the impurity operators.

Using Eq.~\eqref{eqn:H}, we can write down the Matsubara-frequency action:
\begin{align}
	S &= 
	\sum_{q\in \mathrm{BZ}}\sum_{\omega_n} \bar{\gamma}_{q,\omega_n} \left[-\Gamma^{-1}_q\left(i\omega_n + \mu \right) \right] \gamma_{q,\omega_n} 
	\nonumber
	\\
	&+ \sum_{\omega_n}
	\begin{pmatrix}
		\bar{\phi}_{\omega_n}\,, \bar{\psi}_{\omega_n}
	\end{pmatrix}
	\left[-\mathbf{1}G_0^{-1}\left(i\omega_n + \mu\right)\right]
	\begin{pmatrix}
		\phi_{\omega_n} \\ \psi_{\omega_n}
	\end{pmatrix}
	\nonumber
	\\
	&+ \frac{V}{\sqrt{N}} \sum_{q\in \mathrm{BZ}} \sum_{\omega_n}
	\bar{\gamma}_{q,\omega_n}\left[ \phi_{\omega_n}  +  \psi_{\omega_n} e^{-iqdl} \right]
	\nonumber
	\\
	&+ \frac{V}{\sqrt{N}} \sum_{q\in \mathrm{BZ}} \sum_{\omega_n}
	 \left[\bar{\phi}_{\omega_n} + \bar{\psi}_{\omega_n}  e^{iqdl}\right] \gamma_{q,\omega_n}\,.
	 \label{eqn:S}
\end{align}
By comparing $S$ to $H$, one can deduce that $\gamma$ fields correspond to the states in the atomic chain, and $\phi$ and $\psi$ are $a$ and $b$ impurities, respectively. $\Gamma^{-1}_q\left(i\omega_n + \mu \right) = i\omega_n + \mu -\mathcal{E}_q$ is the inverse Green's function for the adsorbate-free atomic chain, while $G_0^{-1}\left(i\omega_n + \mu\right) = i\omega_n + \mu - \epsilon$ is the inverse Green's function for an isolated adatom.

Action $S$ is used to calculate the partition function $\mathcal{Z} = \int \mathcal{D}\left(\dots\right)e^{-S}$, where $\int \mathcal{D}\left(\dots\right)$ denotes the integration over all the fields. In the following sections, we will employ $\mathcal{Z}$ to explore the impurity-chain coupling and how it leads to the impurity-impurity interaction.

\section{Green's and Spectral Functions}
\label{sec:Greens_and_Spectral_Functions}

When the impurity energy levels hybridize with the bulk, their spectral function becomes broadened. This spectral weight reorganization plays a key role in the impurity interaction. Therefore, we perform a careful analysis of the adatom spectral function starting with the calculation of the impurity Green's functions.

\subsection{Impurity Green's Function}
\label{sec:Impurity Green's Function}

According to Eqs.~\eqref{eqn:H} and \eqref{eqn:S}, the two adsorbates do not interact directly. Instead, they both couple to the atomic chain which performs a mediating role. The interaction can be made explicit using the Green's function obtained from the partition function $\mathcal{Z}$. By integrating out the $\gamma$ (atomic chain) fields in the expression of $\mathcal{Z}$, we obtain
\begin{equation}
	\mathcal{Z} = \mathcal{Z}_\mathrm{C} \int \mathcal{D}\left(\dots\right) e^{
	-\sum_{\omega_n}
	\begin{pmatrix}
		\bar{\phi}_{\omega_n}\,, \bar{\psi}_{\omega_n}
	\end{pmatrix}
	\left[-\mathcal{G}_{\omega_n}^{-1}\right]
	\begin{pmatrix}
		\phi_{\omega_n} \\ \psi_{\omega_n}
	\end{pmatrix}}\,,
	\label{eqn:Z}
\end{equation}
where $\mathcal{Z}_\mathrm{C}$ is the partition function for the atomic chain without the adsorbates. The integral on the right hand side is the partition function for the two adatoms $\mathcal{Z}_\mathrm{A}$ with the interaction due to the chain included. $\mathcal{G}_{\omega_n}$ is the desired $2\times 2$ Green's function for the coupled impurities whose inverse is
\begin{align}
		\mathcal{G}^{-1}_{\omega_n} &=
		 \left[
		G^{-1}_0\left(i\omega_n +\mu\right) - V^2 \Xi^0\left(i\omega_n+\mu \right)
		\right]\times
		\nonumber \mathbf{1} 
		\\
		&-
		V^2 \Xi^l\left(i\omega_n+\mu\right)\times
		\begin{pmatrix}
			0 & 1
			\\
			1 & 0
		\end{pmatrix}
	\label{eqn:G_inv}
\end{align}
and
\begin{align}
	\Xi^m\left(z\right)& = \frac{1}{N}\sum_{q\in\mathrm{BZ}} e^{\pm iqdm}\Gamma_q\left(z\right)  = \frac{1}{2\pi}\oint d\theta \frac{e^{\pm i\theta m}}{z -\mathcal{E}_{\theta/d}}
	 \nonumber
	 \\
	 &=-\frac{t + gE}{2\left(t + gz\right)^2}\frac{\left(\Lambda-\sqrt{\Lambda-1}\sqrt{\Lambda+1}\right)^{|m|}}{\sqrt{\Lambda-1}\sqrt{\Lambda+1}}
	 \,,
	 \label{eqn:Omega}
	 \\
	 \Lambda &= \frac{E-z}{2\left(t + gz\right)}\,.
	 \label{eqn:Lambda}
\end{align}

To appreciate the physical significance of the $\Xi^m\left(z\right)$ term, notice that it is the Fourier transform of the chain free-particle propagator $\Gamma_q\left(z\right)$. As such, $\Xi^m\left(\mathcal{E} + i0\right)$ is the real-space propagator for a particle with energy $\mathcal{E}$ inside the adatom-free chain. For $\mathcal{E}$ outside the band, $|\Lambda|>1$ so that $\Xi^m\left(\mathcal{E} + i0\right)$ is real and monotonically decaying with $|m|$ because $\Lambda-\sqrt{\Lambda-1}\sqrt{\Lambda+1}<1$. Physically, this is the consequence of there being no propagating states outside the band.

If $\mathcal{E}$ is inside the band, on the other hand, $\Lambda = \cos\left(q_\mathcal{E}d\right) - i0$ so that $\Xi^m\left(\mathcal{E} + i0\right) = -i\pi D_\mathcal{E} e^{i\left|q_\mathcal{E}d m\right|}$, where $q_\mathcal{E}$ is the momentum of the chain state with energy $\mathcal{E}$, and $D_\mathcal{E}$ is the density of states. Hence, the propagation by distance $dm$ is given by the plane wave phase multiplied by the density of states at $\mathcal{E}$. The amplitude of the propagation does not decay with distance as it does in higher dimensions. Since the propagator functions as the coupling term between the two impurities (see Eq.~\eqref{eqn:G_inv}) the magnitude of the interaction between the adatoms at a given energy does not diminish with increased separation.

In the single-impurity case, the retarded Green's function is given by the diagonal terms of Eq.~\eqref{eqn:G_inv} with $i\omega_n + \mu\rightarrow \omega + i0$ so that $\mathcal{G}_{\mathcal{E}}= \left[G^{-1}_0\left(\omega+i0\right) - V^2 \Xi^0\left(\omega +i0 \right) \right]^{-1}$. For $\omega$ within the band, this is a standard single-state Green's function with frequency-dependent broadening due to the imaginary self-energy correction $iV^2 D_\omega$. Outside the band, $\mathcal{G}_{\omega}$ has two poles: one above and one below the band, corresponding to two localized states. Because these two states are non-propagating, they play a minor role in the impurity interaction. Moreover, their spectral weight decreases sharply as $\epsilon$ is moved away from band edges. Therefore, we will neglect them in our qualitative discussion. In the following section, we explore how interaction between two states with simple broadening gives rise to a highly non-trivial spectral function.

\subsection{Spectral Function}
\label{sec:Spectral_Function}

In the two-impurity case, the inverse of the retarded Green's function for $\omega$ inside the band, in accordance with Eq.~\eqref{eqn:G_inv}, is
\begin{equation}
		\mathcal{G}^{-1}_{\omega} =
		\left(\omega - \epsilon\right)\mathbf{1}
		+
		i\pi V^2  D_\omega
		\begin{pmatrix}
		1 &e^{i\left|q_\omega d l\right|}
		\\
		e^{i\left|q_\omega d l\right|}
		& 1
		\end{pmatrix}\,.
	\label{eqn:G_inv_real_omega}
\end{equation}
As discussed above, the diagonal term $i\pi V^2 D_\omega$ describes the coupling between a single impurity and the chain. The consequence of this is that the impurity state becomes distributed over a range of composite impurity-chain states of energy $\omega$. The off-diagonal term in Eq.~\eqref{eqn:G_inv_real_omega} is the coupling between these composite states for the two impurities. This interaction lifts the degeneracy between the mixed impurity-chain states for both adatoms at each $\omega$ and results in two split energy levels. This mixed-level splitting lies at the heart of the impurity interaction.

It is illustrative to separate the real and imaginary parts of Eq.~\eqref{eqn:G_inv_real_omega}:
\begin{align}
		\mathcal{G}^{-1}_{\omega} &=
		\begin{pmatrix}
			\omega - \epsilon & - \pi V^2 D_\omega \sin \left|q_\omega d l\right|
			\\
			- \pi V^2 D_\omega \sin \left|q_\omega d l\right| & \omega - \epsilon
		\end{pmatrix}
		\nonumber
		\\
		&+
		i\pi V^2 D_\omega 
		\begin{pmatrix}
			1&  \cos \left|q_\omega d l\right|
			\\
			\cos \left|q_\omega d l\right| & 1
		\end{pmatrix}\,.
	\label{eqn:G_inv_real_omega_2}
\end{align}
Focusing on the real part, we see that the adatom-adatom coupling splits the hybridized impurity-chain levels by $\sim \pi V^2 D_\omega \sin \left|q_\omega d l\right|$. The oscillatory nature of the coupling term causes the split levels originating from the hybridized states at different $\omega$'s to periodically bunch up in energy.

To explore this periodic splitting, let us treat the density of states $D_\omega$ as approximately constant for some range of $\omega$. With this, for a given value of $l$, the amount of level splitting at each $\omega$ is dictated by the argument of the sine function in Eq.~\eqref{eqn:G_inv_real_omega_2}. Let $W$ be the energy range of non-split states for which one of the split states ends up at the same energy, forming a spectral peak. Very crudely, $W$ can be approximated from $\left(q_{\omega+\frac{W}{2}} - q_{\omega-\frac{W}{2}}\right)dl \sim \pi$. If $W$ is much smaller than the band width of the chain, we get $W \sim \pi / \left(q'_\omega dl\right)$. Hence, the spacing between the energy ranges with high density of split states is inversely proportional to $l$ with the amount of splitting not affected by the separation. In addition, flatter bands yield smaller $W$ due to the $q'_\omega$ term in the denominator.

\begin{figure}
	\includegraphics[width = 3.0in]{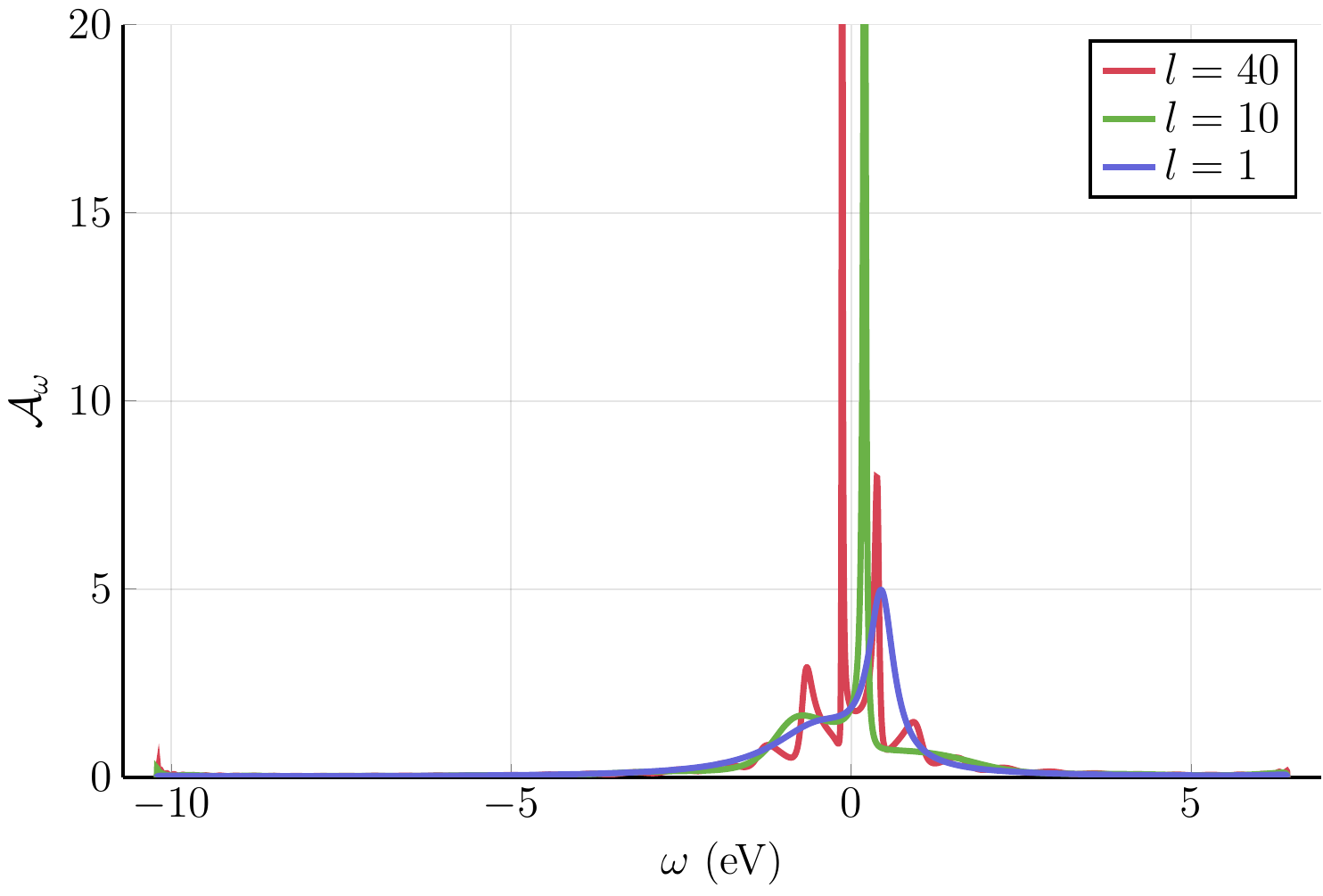}
	\includegraphics[width = 3.0in]{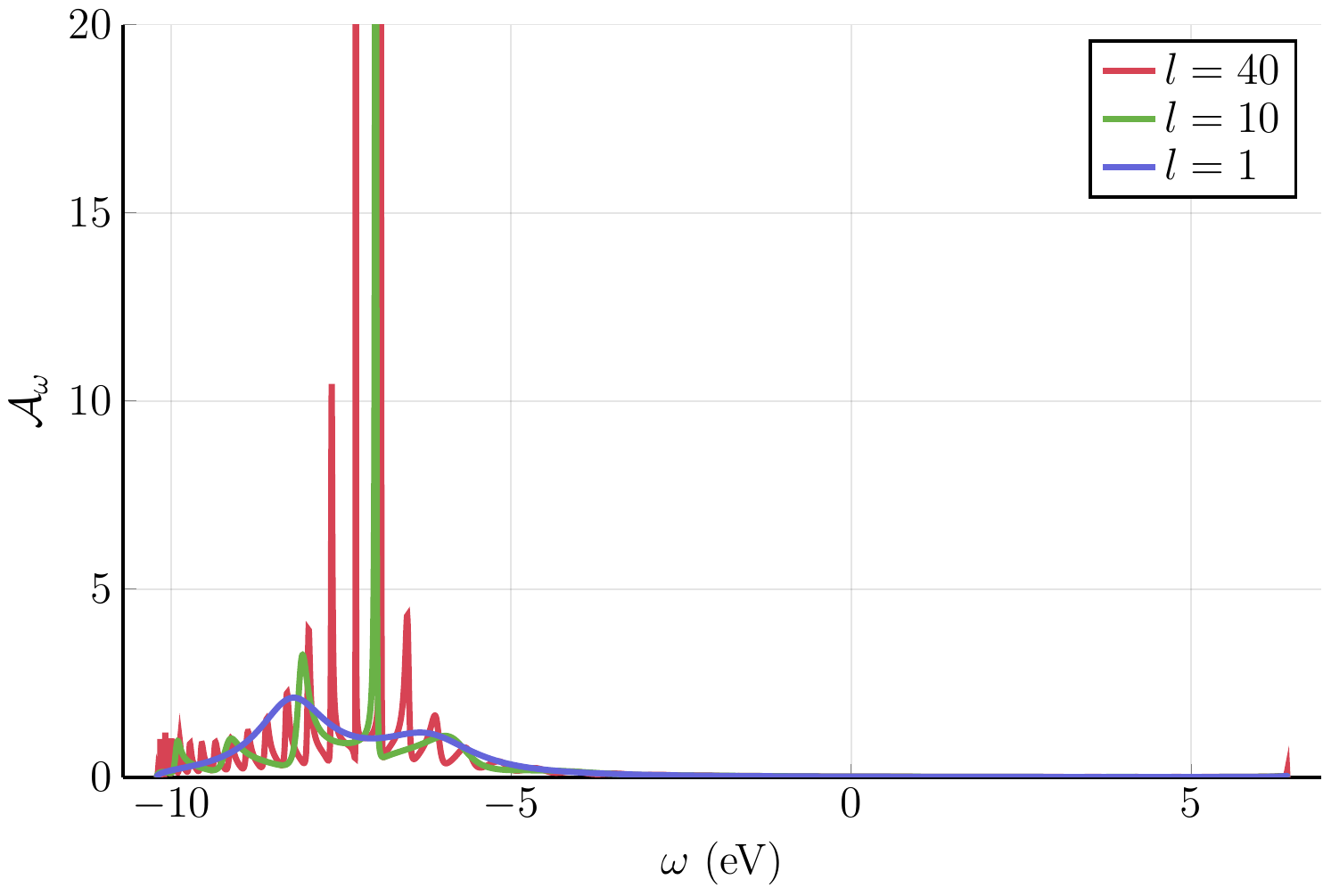}
	\caption{Spectral function for one of the coupled adatoms at $V = t/2$ for several values of $l$. The chain parameters are the same as in Fig.~\ref{fig:Chain}. \textbf{Top:} $\epsilon = 0$ eV. \textbf{Bottom:} $\epsilon = -7.2$ eV, corresponding to a hydrogen atom. This quantity was obtained by taking the difference of the first ionization energies of hydrogen and carbon, and subtracting this value from $E$. The oscillation period becomes smaller at the bottom of the band, as expected.}
	\label{fig:Spectral_Double}
\end{figure}

In order to visualize this periodic concentration of states, we plot the spectral function $\mathcal{A}_\omega = -2\,\mathrm{Im}\left[\mathcal{G}^{11}_\omega\right]$, where
\begin{equation}
	\mathcal{G}^{11}_\omega = \frac{G_0^{-1}\left(\omega\right) - V^2 \Xi^0\left(\omega\right)}{\left[G_0^{-1}\left(\omega \right) - V^2 \Xi^0\left(\omega \right)\right]^2-\left[V^2 \Xi^l\left(\omega \right)\right]^2}
	\label{eqn:G11}
\end{equation}
is the diagonal element of the Green's function in Eq.~\eqref{eqn:G_inv} with $i\omega_n + \mu\rightarrow \omega + i0$, see Fig.~\ref{fig:Spectral_Double}. The spectral function remains centered around $\epsilon$ and displays oscillations whose period decreases with the impurity separation, in agreement with our discussion.

\section{Adatom Interaction Energy}
\label{sec:Adatom_Interaction_Energy}

With the qualitative picture of the two-impurity coupling established, we move to their interaction energy. The first step is to integrate Eq.~\eqref{eqn:Z} over the remaining fields to obtain
\begin{align}
	\mathcal{Z}_\mathrm{A}& =  \mathcal{Z}^+\times \mathcal{Z}^- \,,
	\nonumber
	\\
	\mathcal{Z}^j &=\prod_n \beta \left[G^{-1}\left(i\omega_n + \mu\right) - j V^2 \Xi^l\left(i\omega_n + \mu \right)\right]\,,
	\label{eqn:Z_A}
\end{align}
where $G^{-1}\left(i\omega_n + \mu\right)$ is the diagonal term in Eq.~\eqref{eqn:G_inv}. $\mathcal{Z}^{\pm}$ are the partition functions for the interaction-split impurity-chain states.  From this, one can extract the free energy for the interacting adatoms by recalling that $F = -\left(\ln \mathcal{Z}\right)  / \beta$:
\begin{align}
	&F_\mathrm{A} 
	=
 -\frac{2}{\beta}\sum_{\omega_n}
	\ln\left[\beta \left(i\omega_n + \mu - \epsilon - V^2  \Xi^0\left(i\omega_n+\mu\right) \right)\right] 
	\nonumber
	\\
	&-
	\frac{1}{\beta}\sum_{\omega_n}\ln\left[1 - \left(\frac{V^2 \Xi^l\left(i\omega_n + \mu\right)}{i\omega_n + \mu - \epsilon - V^2 \Xi^0\left(i\omega_n+\mu \right)}\right)^2 \right]\,.
	\label{eqn:F_A}
\end{align}
The first term is the energy of the non-interacting adatoms, coupled to the atomic chain. By subtracting it from $F_A$, we are left with the second term, which we identify as the interaction energy between the adatoms. The Matsubara frequency summation in Eq.~\eqref{eqn:F_A} is easiest to perform at $T = 0$, where it can be replaced by an integral:
\begin{equation}
	F_I = - \int \frac{d\omega}{2\pi} \ln\left[1 - \left(\frac{V^2 \Xi^l\left(i\omega + \mu\right)}{i\omega + \mu - \epsilon - V^2 \Xi^0\left(i\omega+\mu \right)}\right)^2 \right]\,.
	\label{eqn:F_I}
\end{equation}

Before computing the interaction energy from Eq.~\eqref{eqn:F_I}, let us perform a qualitative analysis using Eq.~\eqref{eqn:G_inv_real_omega_2}. At $T = 0$, the energy of the system is calculated by summing the energies of the filled state. For the states located far from the Fermi level, both split and non-split hybridized states are either filled or empty regardless of the separation $l$. Neglecting the finer point of variable spectral weight, this means that they do not contribute to the interaction energy as $l$ (and, therefore, the amount of splitting) is changed. If the split states are close to the Fermi level, however, changing $l$ can cause them to become filled or empty, modifying the energy of the system. Because $l$ enters inside a periodic trigonometric function, the total energy is periodic in $l$. Finally, since the interaction energy is defined as the difference between the coupled and independent-impurity energies, it exhibits the same $l$-periodicity.

From Eq.~\eqref{eqn:G_inv_real_omega_2}, one can see that the wavenumber of these energy oscillations is $2q_{\mu}d$, since only the levels close to the Fermi level contribute. The factor of $2$ comes from the fact that the two split levels move in opposite directions as $l$ varies with wavenumber $k_Fd = q_{\mu}d$. From this, for $k_F d < \pi/2$, the period of oscillations is $\pi /\left(k_Fd\right)$, reminiscent of the Friedel oscillations. At $k_Fd = \pi/2$, the oscillatory period is 2 so that the maxima and minima of $F_I$ alternate with every $l$. If $k_Fd$ exceeds $\pi / 2$, the period becomes less than $2$. In this case, the energy oscillation wavenumber is outside its first Brillouin zone since the period is smaller than lattice spacing. Shifting the wavenumber back to the first BZ gives the period $\pi / \left(\pi - k_Fd\right)$.
\begin{figure}[t]
	\includegraphics[width = 3in]{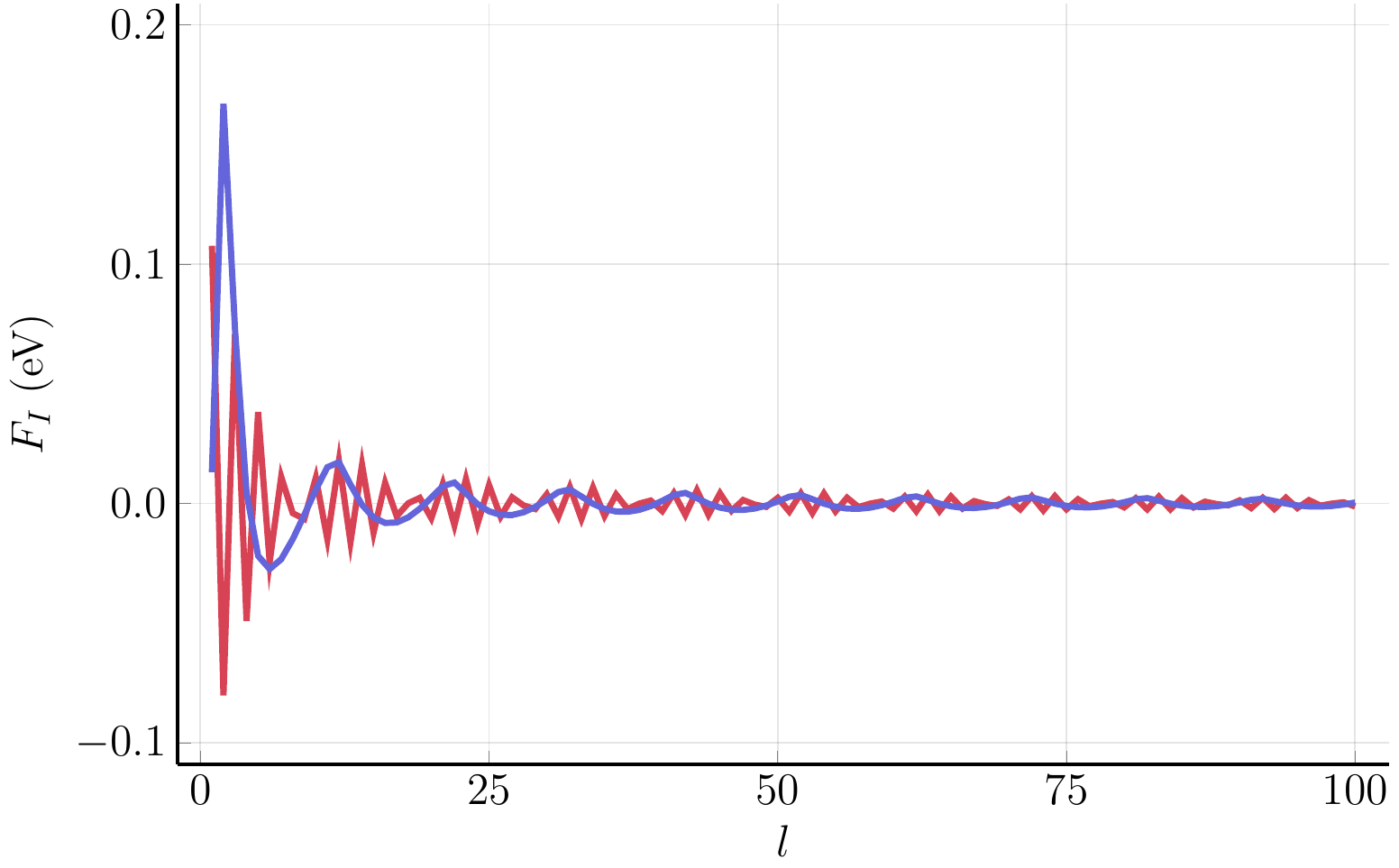}
	\includegraphics[width = 3in]{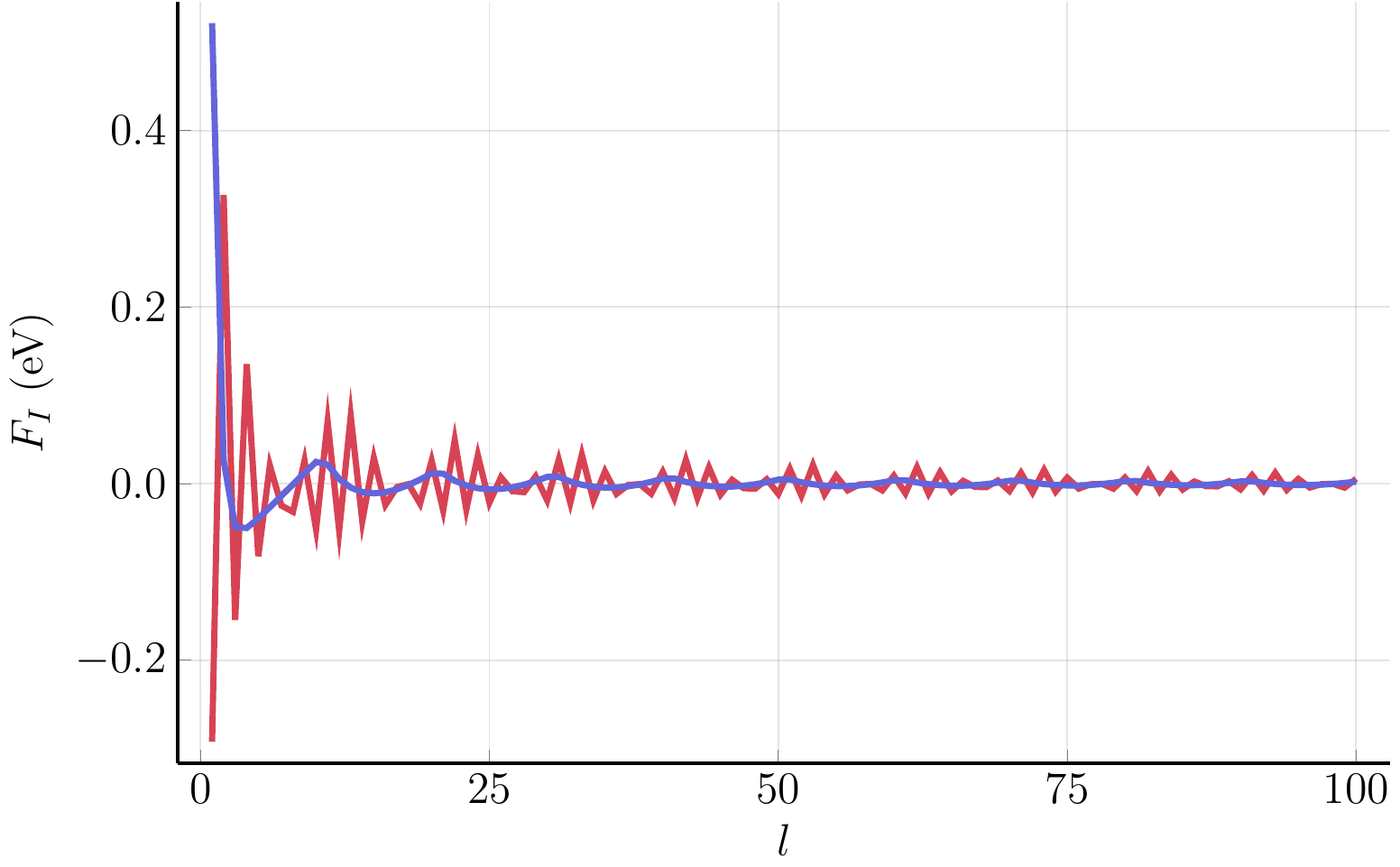}
	\caption{$F_I$ as a function of $l$ for $V = t = 4.49$ eV. The chain parameters are the same as in Figs.~\ref{fig:Chain} and \ref{fig:Spectral_Double}. \textbf{Top:} $\epsilon = 0$. \textbf{Bottom:} $\epsilon = -7.2$. The smooth curve corresponds to $\mu = -10$ eV, while the jagged has $\mu = -5.93$ eV. These $\mu$'s correspond to the values of $k_F$ chosen to give smooth or beat-like $F_I$ oscillations with period equal to 10 (see main text for details).}
	\label{fig:Interaction}
\end{figure}

If the period of the $F_I$ oscillations is much larger than the interatomic separation in the chain, the interaction energy undergoes a smooth variation with $l$. As the period approaches 2 ($k_F d \approx \pi / 2$), the granularity of $l$ prevents the smooth oscillations of $F_I$. Instead, they acquire the form of beats with the period $\pi / \left|\pi - 2k_F d\right|$. This can be visualized by plotting a sine curve on a lattice where the period of the function is slightly larger or smaller than twice the lattice constant. With this in mind, we plot the interaction as a function of $l$ in Fig.~\ref{fig:Interaction}. To illustrate the smooth variation and the beat-like pattern of the interaction energy, we choose two values of $\mu$ for which the smooth period and the beat length is 10 lattice constants. These $\mu$'s are obtained by solving for $k_F$ using the expressions above and then calculating the chemical potential from Eq.~\eqref{eqn:Dispersion}.
 
One can see that the amplitude of energy oscillations decreases with increasing $l$. This does not contradict the point we made earlier about the magnitude of the propagator not changing with the separation. The amplitude of the energy oscillations is related to the number of states that get filled or emptied with varying $l$. This can be understood as follows. When talking about the peaks of the two-impurity spectral function, we introduced a quantity $W$. This quantity describes the energy range of non-split states that contribute to a spectral peak and can also be thought of as the spacing between the peaks. In other words, larger $W$ results in more states in a peak and, therefore, larger energy oscillation as these are filled and emptied with varying $l$. As was shown above, $W$ decreases as $l$ gets bigger, reducing the amplitude of the energy oscillations.

\begin{figure}
	\includegraphics[width = 3in]{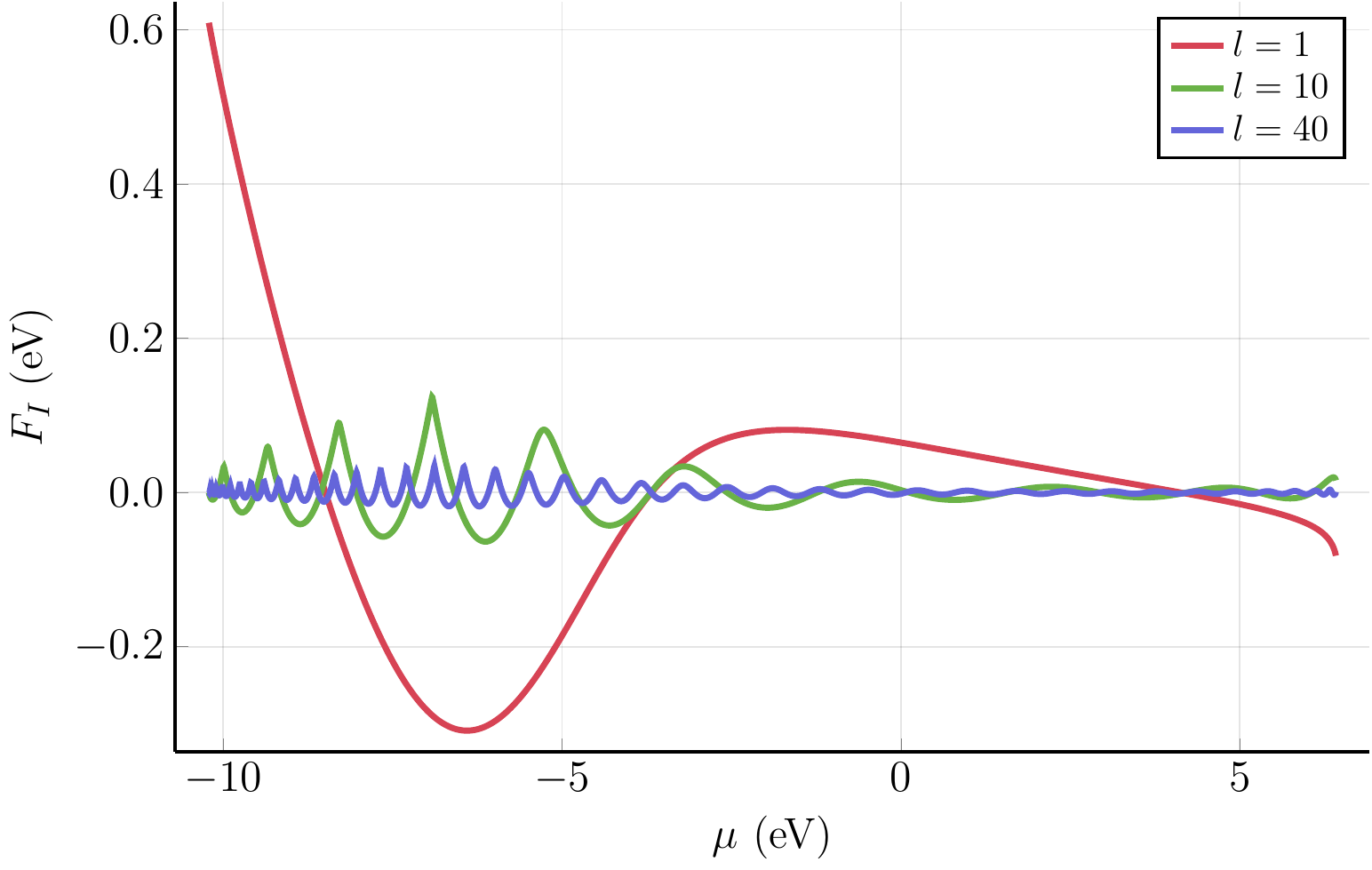}
	\caption{$F_I$ as a function of $\mu$ for different $l$'s at $\epsilon = -7.2$ eV and $V = t$. Larger separations yield more oscillatory behavior.}
	\label{fig:Interaction_mu}
\end{figure}

In addition to the smaller amplitude of energy oscillations, the reduction of peak spacing with increased $l$ also makes the system more sensitive to changes in $\mu$ (see Fig.~\ref{fig:Interaction_mu}). For large $W$, small variation in $\mu$ keeps the Fermi level within the same peak of the spectral function. As $W$ is reduced, changing $\mu$ causes it to cross multiple spectral peaks, leading to oscillations of the interaction energy.

Finally, we address the role of the adatom energy level $\epsilon$. Because of the broadening, its position relative to $\mu$ determines the number of states that participate in the energy oscillations. Thus, having it closer to $\mu$ leads to a larger amplitude, as seen in Fig.~\ref{fig:Interaction}.

In order to validate the theoretical model, as well as to gain insight into how the adsorbates interact in a concrete case, we create a model 1D C-chain with 2 H adsorbates, and determine $F_I$ \emph{ab initio}. We note that, because of the non-decaying propagator, an infinite (periodic) chain would result in each H adsorbate interacting with an infinite number of other H adsorbates. We therefore choose to model the system as an isolated 36 C-atom fragment, with one H adsorbate fixed to an edge while the location of the second is varied along the chain. The interaction energy of the impurities can be calculated as
\begin{equation}
	F_I = E_{b}^{H_{0}H_{l}} - E_{b}^{H_0} - E_{b}^{H_l},
	\label{eqn:F_I_dft}
\end{equation}
where $E_{b}^{H_{0}H_{l}}$ is the binding energy of the two-impurity system with H adsorbates on C atoms at positions $0$ and $l$, while $E_{b}^{H_0}$ and $E_{b}^{H_l}$ represent the binding energies of one-impurity systems with the H adsorbate at position $0$ and $l$, respectively. 

\begin{figure}[h]
	\includegraphics[width = 3in]{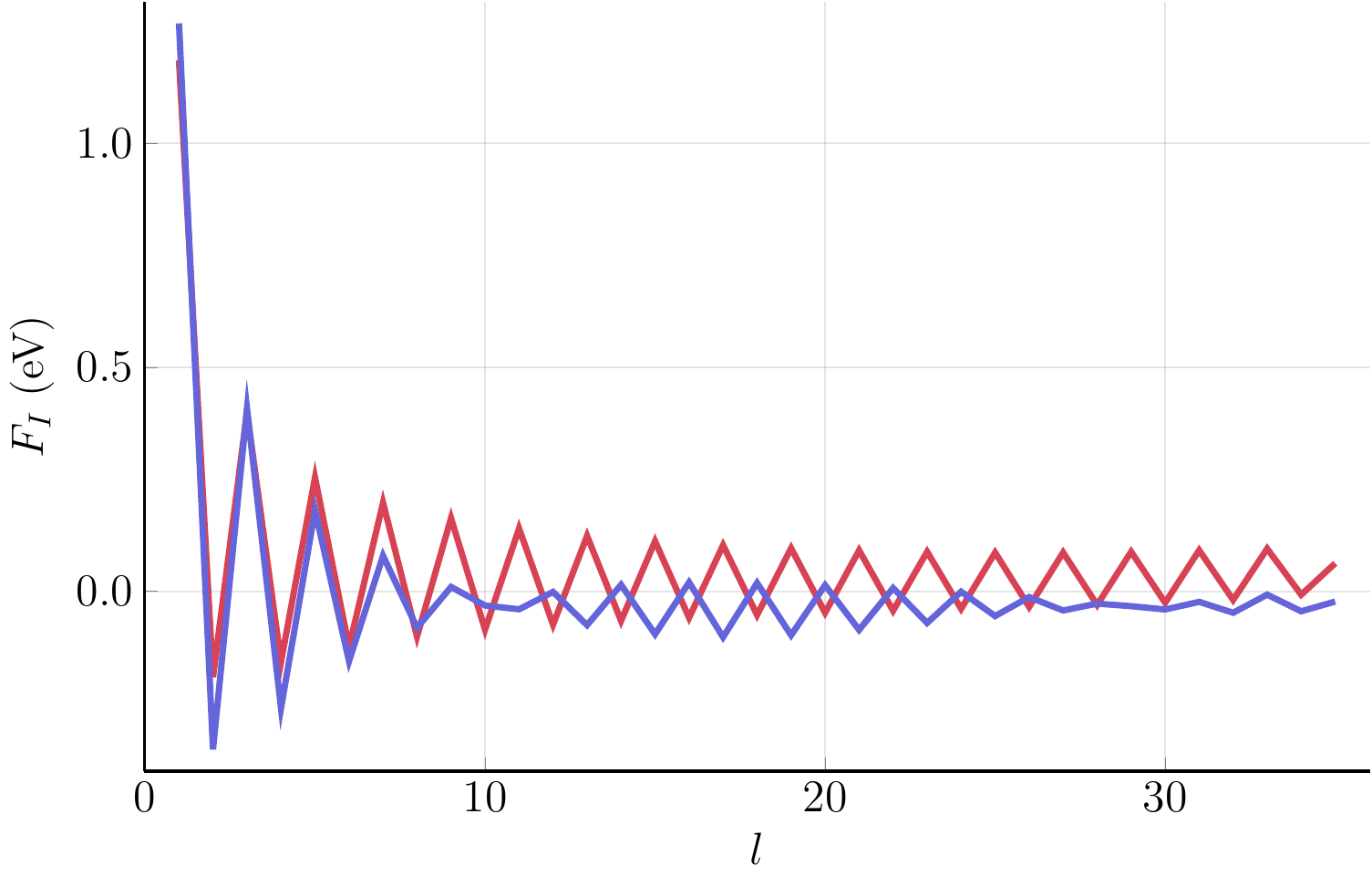}
	\includegraphics[width = 3in]{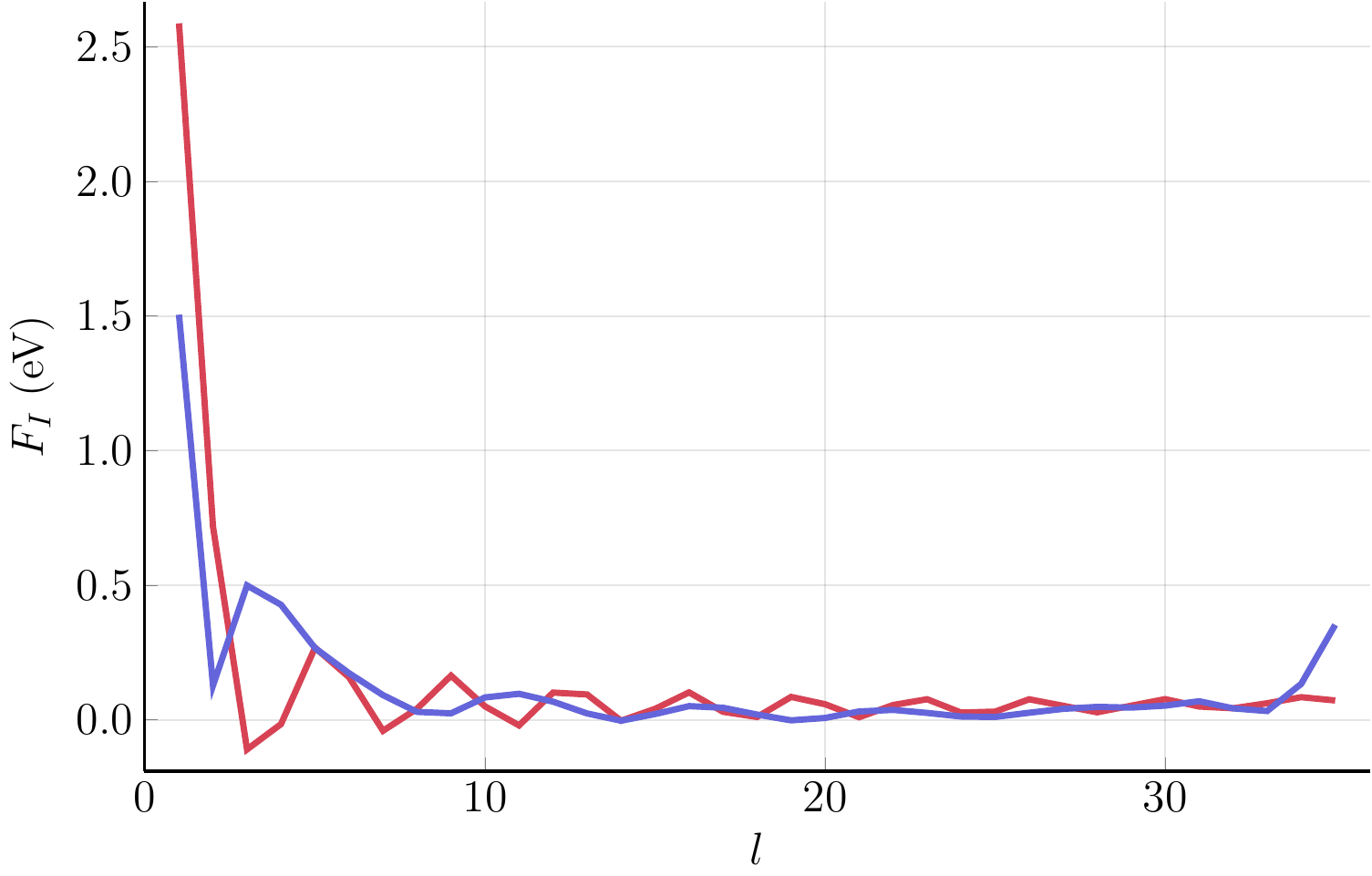}
	\caption{$F_I$ as a function of $l$ obtained from DFT calculations. \textbf{Top:} Results at (red) and close to (blue) half filling. For the blue curve, 5 electrons were removed from the 36C + 2H system. Shifting the Fermi level from half filling creates the beat-like pattern. \textbf{Bottom:} Results for $\mu$ close to the bottom of the band. For the red (blue) curve, 36 (54) electrons were removed from the system. The smaller $k_F$ of the blue curve gives rise to a smoother $F_I$ profile. The sharper features of the red curve are a consequence of its shorter period of oscillations.}
	\label{fig:Interaction:DFT}
\end{figure}

Fig.~\ref{fig:Interaction:DFT} shows $F_I$ as a function of $l$ for four doping levels. Exactly at half filling, the value of $F_I$ oscillates with a period equal to $2$, while its amplitude decreases with $l$, in agreement with the discussion above. As $\mu$ is lowered, the period becomes slightly larger than 2, leading to the formation of a beat pattern, consistent with the picture shown in Fig.~\ref{fig:Interaction}. When $\mu$ is moved close to the bottom of the $\pi$ bands, $k_F$ decreases sufficiently to yield the period of energy oscillations much larger than the atomic spacing, leading to a smoother variation with $l$.

We observe a strong qualitative agreement between the DFT and analytical results, as seen from Figs.~\ref{fig:Interaction} and ~\ref{fig:Interaction:DFT}. From the quantitative standpoint, the results are of the same order of magnitude. The differences in the energy scales can be attributed to two main factors. First, the hopping parameter $V$ used in the analytical calculations was not calculated exactly. Instead, we chose a representative value with the right order of magnitude when compared to the C-C hopping element. In addition, the DFT results were obtained using a finite chain where one of the adsorbates was hosted by an edge atom. As such, finite-size effects play a role in determining $F_I$.

\section{Conclusion}
\label{sec:Conclusion}

We have performed a detailed study of bulk-mediated interaction between adsorbates in 1D atomic chains. We have shown using analytical and \emph{ab initio} methods that the interaction energy between these adsorbates exhibits an oscillatory behavior which can be tuned by changing the doping of the system. The qualitative understanding gained from studying this simple system can be applied to more complicated configurations. More specifically, our results suggest an effective approach to further exploring adatom interactions in 2D materials where interaction control by gating could potentially be used for adatom manipulation and assembly.

\section{Acknowledgements}
\label{sec:Acknowledgements}
The authors acknowledge the National Research Foundation, Prime Minister Office, Singapore, under its Medium Sized Centre Programme. A. R. acknowledges the support by Yale-NUS College (through grant number R-607-265-380-121). K. N. and S.Y. Q. acknowledge support from Grant NRF-NRFF2013-07 from the National Research Foundation, Singapore. Computations were performed on the NUS CA2DM cluster.

\appendix
\section{Computational Methods}
\label{sec:Methods}
DFT calculations were performed in VASP~\cite{Kresse1996eis} using PBE~\cite{Perdew1997gga} PAW potentials~\cite{Blochl1994paw, Kresse1999fup} and a kinectic energy cutoff of 400 eV for wavefunctions. In the case of the 1D C chain (Fig.~\ref{fig:Chain}), periodic images along the directions orthogonal to the chain axis were separated by 15 \AA\ of vacuum in order to avoid spurious Coulomb interactions. The charge density of the perodic chain was calculated with a uniform Brillouin zone sampling of $36 \times 1 \times 1$ k-points. The corresponding bandstructure was computed at 100 evenly-spaced k-points between each high symmetry point.  In the case of the isolated 36 C-atom fragment, a 15 \AA\ vacuum was also imposed in the direction along the chain. Variations in $\mu$ were achieved by removing electrons from the system while applying a compensating background charge.

Numerical integration and plotting were performed using Julia programming language.~\cite{Bezanson2017} The code can be found at \href{url}{https://github.com/rodin-physics/chain-impurity-interaction}.

\bibliography{Chain_Dimer_Interaction.bib}
\end{document}